\documentclass[twocolumn,pra]{revtex4}
\usepackage{amsmath}
\usepackage{amssymb}
\usepackage{graphicx}
\usepackage{epsfig}
\usepackage{bbm}
\usepackage{mathrsfs}

\newcommand{\be}{\begin{equation}}
\newcommand{\ee}{\end{equation}}
\newcommand{\bey}{\begin{eqnarray}}
\newcommand{\eey}{\end{eqnarray}}
\newcommand{\bw}{\begin{widetext}}
\newcommand{\ew}{\end{widetext}}

\newcommand{\ov}{\overline}

\newcommand{\ra}{\rangle}
\newcommand{\la}{\langle}

\newcommand{\bI}{\boldsymbol{I}}
\newcommand{\bJ}{\boldsymbol{J}}

\newcommand{\bm}{\boldsymbol{m}}

\newcommand{\HH}{{\mathscr{H}}}

\newcommand{\ba}{\begin{array}}
\newcommand{\ea}{\end{array}}
\newcommand{\bi}{\begin{itemize}}
\newcommand{\ei}{\end{itemize}}
\newcommand{\bem}{\begin{enumerate}}
\newcommand{\eem}{\end{enumerate}}

\makeatother

\begin{document}

\title{Convergent perturbation expansion of energy eigenfunctions
 on unperturbed basis states in classically-forbidden regions
 }

 \author{Jiaozi Wang and Wen-ge Wang
 \footnote{ Email address: wgwang@ustc.edu.cn}
  }

\affiliation{Department of Modern Physics, University of Science and Technology of China,
Hefei, 230026, China}

\date{\today}

\begin{abstract}
  We study properties of eigenfunctions of perturbed systems,
  given on the eigenbases of unperturbed, integrable systems.
  For a given pair of perturbed and unperturbed systems,
  with respect to the energy of each perturbed state,
  the unperturbed basis states can be divided into two groups:
  one in the classically-allowed region and the other in the classically-forbidden region;
  correspondingly, the eigenfunction of the perturbed state is also divided into two parts.
  In the semiclassical limit, it is shown that, making use of
  components of the eigenfunction in its classically-allowed region,
  its components in the classically-forbidden region  can be written in the form of
  a convergent perturbation expansion, which is valid for all perturbation strengths.
\end{abstract}

%\pacs{05.45.Mt, 03.65.-w, 05.45.Pq }
% 03.65.-w Quantum mechanics
% 05.45.Mt Quantum chaos; semiclassical methods
% 05.45.-a Nonlinear dynamics and chaos
% 03.65.Aa Quantum systems with finite Hilbert space
% 67.40.Db Quantum statistical theory; ground state, elementary excitations
% 05.45.Pq Numerical simulations of chaotic systems

\maketitle

%\tableofcontents

\section{Introduction}

 Energy eigenfunctions (EFs) play an essential role in our understanding of
 many properties of quantum systems and have been widely studied
 (see, e.g., Refs.\cite{Haake,CC94book,Berry77,Berry91,KpHl,Heller87,Bies01,Sr96,Sr98,SrCF,Backer02,Urb03,Kp05}).
 For systems that have classical counterparts,
 such an EF can be divided into two parts in a basis that possesses a clear semiclassical meaning,
 that is, a part in the classically-allowed region and
 a part in the classically-forbidden region.
 Both parts are of importance in the study of quantum systems.

 In the configuration space, the semiclassical theory supplies a useful tool
 in the study of the parts of EFs in their classically-allowed regions
 \cite{Gutzbook,CC94book,Stockmann,Haake}.
 It predicts, e.g., the spatial correlation function of chaotic EFs
 and leads to Berry's conjecture for chaotic EFs
 \cite{Berry77,KpHl,Heller87,Bies01,Sr96,Sr98,SrCF,Backer02,Urb03,Kp05}.
 Meanwhile, much less is known for those parts of EFs in classically-forbidden regions.
 In some cases, the WKB method is useful.
 And, a formal approach to these parts was given in discretized coordinates,
 which shows that their components can be
 written in the form of a generalized Brillouin-Wigner perturbation expansion (GBWPE),
 as a function of the components at the border separating
 the classically allowed and forbidden regions \cite{cpl04-config,cpl05}.
 This approach was found useful in the study of some special topics such as the tunnelling effect,
 but, generally it may become quite cumbersome in application, particularly when
 resuming the continuity of coordinates.

 Quite often, one is interested in EFs expanded in unperturbed bases.
 In such a basis, each EF can also be divided into two parts,
 one in classically allowed region and the other in classically forbidden region.
 Like in the case of configuration space, the semiclassical theory may be used in the study of the former part;
 for example, for chaotic systems, it predicts a Gaussian shape for the distribution of rescaled components in
 this parts of EFs \cite{pre18-EF-BC}.
 But, an analytical framework is still lacking, which is generically
 valid for the study of perturbed EFs in classically forbidden regions.
 An interesting question is whether the GBWPE,
 which was first introduced in Ref.\cite{pre-98} for EFs in unperturbed bases, 
 could be useful in this study.

 In this paper, we give a positive answer to the above question.
 Specifically, in the case that unperturbed systems are integrable systems,
 we show that in the semiclassical limit the classically-forbidden part of each EF
 can be expanded in a convergent GBWPE.
 Since the GBWPE has already been found useful in explaining some properties of EFs
 \cite{pre-98,pre00,pre02-LMG,pre01-trunc,EFchaos-WW,pre02-WBRM,ctp01-ratio,cpl04-config,cpl05},
 it is reasonable to expect that it may supply a useful method
 in the study of perturbed EFs in their classically-forbidden regions.

 The paper is organised as follows.
 In Sec.\ref{sect-GBWPT}, we discuss basic contents of a semiperturbative theory, which is based on GBWPE.
 In particular, we discuss an important concept in this theory,
 namely, perturbative (PT) parts of EFs, defined as the largest parts of EFs
 that can be expanded in a convergent GBWPE.
 In Sec.{\ref{sect-NPT}},  in the semiclassical limit, we show that
 PT parts of EFs coincide with their classically-forbidden parts.
 Numerical tests of the above prediction are given in Sec.\ref{app-models}.
 Finally, conclusions and discussions are given in Sec.\ref{sect-Conclusion}.

\section{Semiperturbative theory}
\label{sect-GBWPT}

 In Sec.\ref{sect-orgin-GBWPT}, we recall a basic form of the GBWPE,
 which was introduced in Ref.~\cite{pre-98}.
 Then, in Sec.\ref{sect-gene-GBWPT}, we give a generic form of the GBWPE.
 In Sec.\ref{sect-semiperturb}, we discuss a framework, suggested by the GBWPE,
 for the study of structural properties of EFs.

\subsection{A basic form of GBWPE}\label{sect-orgin-GBWPT}

 Consider a perturbed system with a Hamiltonian written as
\begin{equation}\label{H}
 H=H_{0}+\lambda V,
\end{equation}
 where $H_{0}$ indicates a generic unperturbed Hamiltonian and $\lambda V$ represents a generic perturbation
 with a running parameter $\lambda$.
 For the sake of convenience in discussion,
 we assume that the system $H$ has a finite Hilbert space $\HH$, with a dimension denoted by $N_d$.
 Eigenstates of $H$ and of $H_{0}$ are denoted by $|\alpha\rangle$ and $|k\rangle$,
 respectively,
\begin{subequations}\label{Seq}
\begin{gather}\label{Seq-H}
 H|\alpha\rangle=E_\alpha|\alpha\rangle,
 \\ H_{0}|k\rangle=E_{k}^{0}|k\rangle. \label{Seq-H0}
\end{gather}
\end{subequations}
 The label $\alpha$ is in the order of increase of energy, while, there is no generic restriction to the order of $k$.
 Components of the EFs are denoted by $C_{\alpha k} = \la k|\alpha\ra$.

 For the sake of simplicity in discussion,
 we assume that $E^0_k \ne E_\alpha$ for all the unperturbed and perturbed energies.
 Furthermore, we assume that the perturbation $V$ has a zero diagonal term in the eigenbasis of $H_0$,
 that is, $V_{kk}=0$ for all $|k\ra$,  where $V_{kk}=\langle k |V|k\rangle$.
 In fact, in the case that $V_{kk} \ne 0$ for some $k$,
 one may consider to employ a new unperturbed Hamiltonian, which is equal to
 $H_0 + \lambda \sum_k  V_{kk} |k\rangle\langle k|$, and a new perturbation,
 equal to $\lambda V -\lambda \sum_k  V_{kk} |k\rangle\langle k| $, such that the total
 perturbed Hamiltonian remains unchanged.

 In order to obtain a GBWPE, one may focus on one perturbed state $|\alpha\rangle$
 and divide the whole set of unperturbed states $|k\rangle$ into two subsets,
 denoted by $S$ and $\ov S$.
 We use $\HH_S$ and $\HH_{\ov S}$ to denote the subspaces of $\HH$,
 which are spanned by $|k\ra \in S$ and $|k\ra \in \ov S$, respectively.
 Projection operators on these two subspaces are denoted by $P_S$ and $Q_{\ov S}$, respectively,
\begin{equation} \label{PQ}
P_{S}=\sum\limits _{|k\rangle\in S}|k\rangle\langle{k}|, \quad
\ Q_{\ov S} =\sum\limits _{|k\rangle \in {\ov S}}|k\rangle\langle k|.
\end{equation}
 Clearly, $P_S + Q_{\ov S} =1$ and $P_S H_0 = H_0 P_S$.
 These two projection operators divide the perturbed state $|\alpha\ra$ into two parts,
\begin{gather}\label{eq-apq}
 |\alpha \ra = |\alpha_P\ra + |\alpha_Q\ra,
\end{gather}
 where $|\alpha_{P}\rangle\equiv{P_{{S}}|\alpha\rangle}$ and
 $|\alpha_{Q}\rangle\equiv Q_{{\ov S}}|\alpha\rangle$.

 Multiplying both sides of Eq.(\ref{Seq-H}) by $P_S$, one gets that
 $(E_\alpha -H_0)|\alpha_P\ra = \lambda  P_S V|\alpha\ra$.
 This gives that
\begin{gather}\label{alpha-P-1}
|\alpha_{P}\rangle=T|\alpha\rangle,
\end{gather}
 where $T$ is defined by
\begin{equation}\label{T-alpha}
 T :=  \frac{1}{E_\alpha-H_{0}}\lambda P_S {V}.
\end{equation}
 Substituting Eq.(\ref{eq-apq}) into the right-hand side (rhs) of Eq.(\ref{alpha-P-1})
 and noting the fact that $T=P_S T$ and $|\alpha_P\rangle = P_S |\alpha_P\rangle$,
 the part $|\alpha_P \ra$ can be written as
\be\label{alpha-T}
|\alpha_{P}\rangle=T|\alpha_{Q}\rangle+W_S|\alpha_{P}\rangle,
\ee
where $W_S$ is an operator acting on the subspace $\HH_S$, defined by
\be\label{U}
W_S := P_S T P_S.
\ee
 From Eq.(\ref{alpha-T}), one gets the following iteration expansion,
\begin{align}\label{alphaP-W}
|\alpha_{P}\rangle &  = \sum_{k=1}^{n}(W_S)^{k-1}T|\alpha_{Q}\rangle + (W_S)^{n}|\alpha_{P}\rangle,
\end{align}
 or, equivalently,
\begin{align}\label{alphaP-TQP}
|\alpha_{P}\rangle & =\sum_{k=1}^{n}T^{k}|\alpha_{Q}\rangle + (W_S)^{n}|\alpha_{P}\rangle.
\end{align}

 Equation (\ref{alphaP-TQP}) shows that, if the following condition,
\begin{equation}
 \lim _{n \to \infty }  \langle \alpha_P |(W_S^{\dagger})^n
 (W_S)^n |\alpha_P  \rangle =0,  \label{conv}
\end{equation}
is satisfied,  then, $|\alpha_{P}\rangle$ can be expanded
 as a convergent perturbation expansion given below,
\begin{equation}\label{alpha-ovs}
|\alpha_{P}\rangle=T|\alpha_{Q}\rangle+T^{2}|\alpha_{Q}\rangle
 +T^{3}|\alpha_{Q}\rangle+\cdots.
\end{equation}
 The rhs of Eq.(\ref{alpha-ovs}) is called a \emph{GBWPE} \cite{pre-98}.
 Note that  Eq.(\ref{alpha-ovs}) is valid for all perturbation strengths, while the size of the part $|\alpha_P\ra $ is $\lambda$-dependent.

 We consider the generic case, in which
 validity of the condition (\ref{conv}) does not depend on concrete properties of $|\alpha_P\ra$.
 In this generic case, this condition is written as
\begin{gather}\label{conv-W}
 \lim_{n\to \infty}  (W_S) ^n =0.
\end{gather}
 We use $|\nu\rangle $ and $w_{\nu }$ to denote the eigenvectors
 and eigenvalues of $W_S$,
\begin{gather}\label{Tn-WS}
 W_S |\nu\rangle = w_{\nu } |\nu\rangle .
\end{gather}
 Note that the operator $W_S$ is not Hermitian and, hence, its eigenvectors $|\nu\ra$ are not necessarily
 orthogonal to each other.
 Due to the finiteness of the Hilbert space and the fact that $E_k^0 \ne E_\alpha$,
 all the values of $w_\nu$ are finite.
 It is not difficult to see that the condition (\ref{conv-W}) is equivalent to the following one,
\begin{equation} \label{inequ}
 | w_{\nu }| < 1 \quad \forall \ |\nu\rangle.
\end{equation}
 To summarize, when Eq.(\ref{inequ}) is satisfied, the part $|\alpha_P\ra$ has a
 convergent perturbation expansion in the form of Eq.(\ref{alpha-ovs}).

\subsection{A generic form of GBWPE}\label{sect-gene-GBWPT}

 In this section, we present a form of the GBWPE that is more generic than that
 discussed in the previous section.
 To this end, let us write the operator $W_S$ on the rhs of Eq.(\ref{alpha-T}) as $(W_S-a+a)$,
 where $a$ is a parameter, and move the part $a|\alpha_P\ra$ to the
 left-hand side of the equation.
 Then, simple derivation gives that
\begin{gather}\label{alpha-TK}
|\alpha_{P}\rangle=\frac{T}{1-a}|{\alpha}_{Q}\rangle+W_{a}|\alpha_{P}\rangle,
\end{gather}
 where  $W_a$ is defined by
\begin{gather}\label{eq-WK}
 W_a := \frac{P_S(W_S-a)P_S}{1-a}.
\end{gather}
Noting that
\be
T|\alpha_{Q}\rangle=P_{S}T|\alpha_{Q}\rangle=P_{S}(T-a)|\alpha_{Q}\rangle,
\ee
 $|\alpha_P\rangle$ can be written as
\be
|\alpha_{P}\rangle=T_{a}|\alpha_{Q}\rangle+W_{a}|\alpha_{P}\rangle,
\ee
where
\be
T_{a} :=\frac{P_{S}(T-a)}{1-a}.
\ee
Obviously,
\be
W_a = P_S T_a P_S.
\ee

 For the sake of convenience in later discussions, we introduce an operator $A_n$,
\be \label{An}
A_n := (W_a^\dagger)^n (W_a)^n.
\ee
Then, following arguments similar to those given in the previous section, one finds that,
 when the following condition,
\begin{equation}\label{eq-An}
 \lim _{n \to \infty }  \langle \alpha_P |A_n|\alpha_P  \rangle =0,
\end{equation}
 is satisfied, $|\alpha_P\ra$ has the following convergent perturbation expansion,
\begin{equation}\label{alpha-ovs-K}
 |\alpha_P\ra =T_a|\alpha_{Q}\rangle+T_a^{2}|\alpha_{Q}\rangle
 +T_a^{3}|\alpha_{Q}\rangle+\cdots.
\end{equation}
 This is a generic form of the GBWPE.
 In a generic case, the condition (\ref{eq-An}) has an $|\alpha_P\ra$-independent form, written as
\be\label{conditionWsE}
\lim_{n\rightarrow \infty}A^\psi _n =0 \quad  \forall|\psi\rangle\in\mathscr{H}_{{S}},
\ee
 where
\begin{gather}\label{An-psi}
 A_n^\psi \equiv \langle \psi | A_n |\psi \rangle = \text{Tr}(\rho_{\psi}A_n).
\end{gather}
 Here, $\rho_\psi = | \psi \ra \la \psi|$.
 Note that, due to the denominator $(1-a)$ in $T_a$, certain rescaling is effectively performed
 in the expansion in Eq.(\ref{alpha-ovs-K}), in comparison with that in Eq.(\ref{alpha-ovs}).

 Let us use $|\nu_a\rangle $ and $w^a_{\nu }$ to denote the eigenvectors
 and eigenvalues of $W_a$,
\begin{gather}\label{Tn-Wa}
 W_a |\nu_a\rangle = w^a_{\nu } |\nu_a\rangle .
\end{gather}
 Then, the condition (\ref{conditionWsE}) can be equivalently written as
\be\label{con-WK}
|w^a_\nu|<1 \quad \forall \ |\nu_a\rangle.
\ee
 Furthermore, noting that  Eq.(\ref{eq-WK}) gives that
\be
w_{\nu}^{a}=\frac{w_{\nu}-a}{1-a},
\ee
 the condition (\ref{con-WK}) has the following equivalent form,
\be\label{con-WK2}
|w_\nu -a|<|1-a| \quad \forall \ |\nu \rangle.
\ee

 The condition (\ref{con-WK2}) can be further simplified.
 To achieve this goal, let us first discuss the specific case that all the values of $w_\nu$ are real.
 We denote the maximum and minimum
 values of $w_\nu$ by $w_ {\rm max}$ and $w_{\rm min}$, respectively.
 Straightforward derivation shows that the condition (\ref{con-WK2}) is satisfied,
 if the parameter $a$ satisfies the following requirements,
\be\label{con-K}
\begin{cases}
a<\frac{1+w_{\rm min}}{2}, & \ \text{if} \ w_{\rm max}<1;\\
a\in\varnothing, & \ \text{if} \  w_{\rm min}<1<w_{\rm max};\\
a>\frac{1+w_{\rm max}}{2} & \ \text{if} \  w_{\rm min}>1;
\end{cases}
\ee
 here, $\varnothing$ represents the empty set.
 Although $W_S$ is not Hermitian, its eigenvalues satisfy the following relation \cite{DMatrix},
 \be
 \sum_\nu w_\nu=\sum_{|k\ra \in S} W_{kk}=\sum_{|k\ra \in S} \frac{V_{kk}}{E_\alpha-E^0 _k},
 \ee
  where $W_{kk} = \la k|W_S|k\ra$.
 Then, since as discussed previously $V_{kk}=0$ for all $|k\ra$, one gets that
 \be\label{sum-nu=0}
 \sum_\nu w_\nu =0.
 \ee
 This implies that $w_{\rm min} < 0$.
 As a result, the first case in the condition (\ref{con-K}) is the only one of relevance.
 Thus, finally, one reach the following conclusion:
 A convergent GBWPE in Eq.(\ref{alpha-ovs-K}) exists,
 if the set $S$ is chosen such that
\be\label{wm-11}
w_{\rm max}<1.
\ee
 Sometimes, a convenient choice of the parameter $a$ is that $a=(w_{\rm max}+w_{\rm min})/{2}$.

\begin{figure}
\includegraphics[width=1\linewidth]{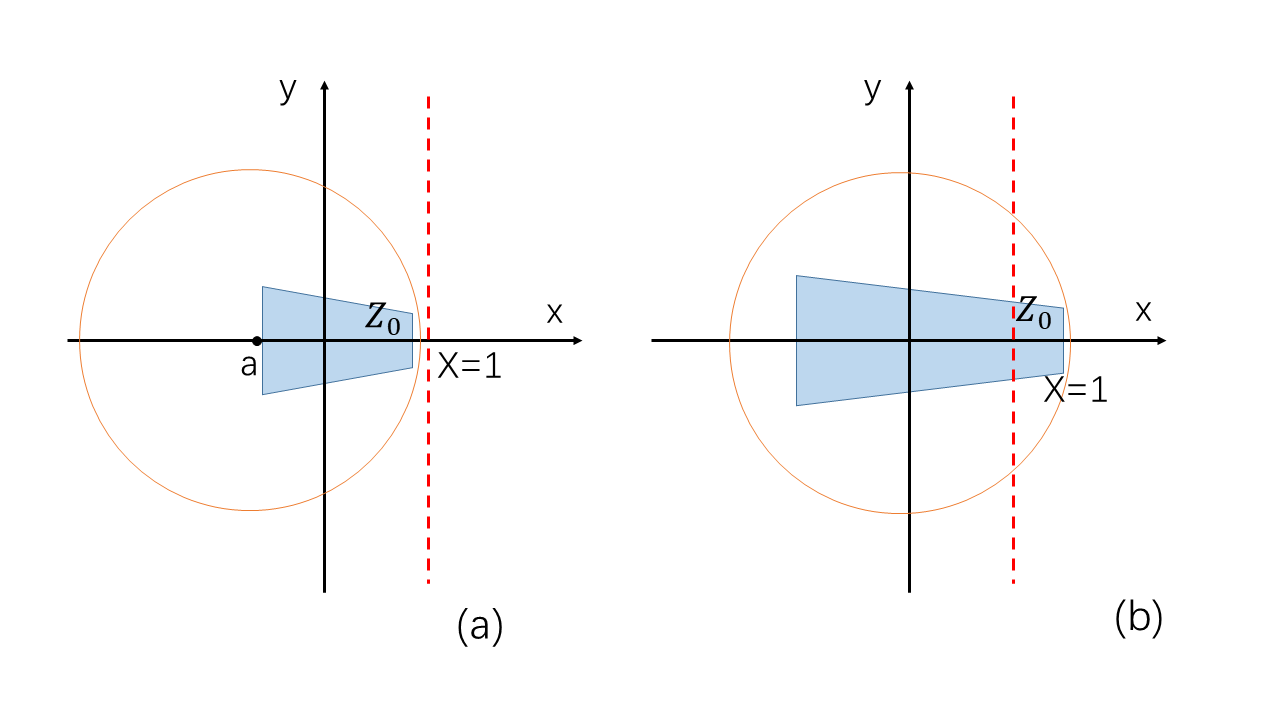}
\caption{ Illustration of situations considered in the derivation of Eq.(\ref{CA}).
}\label{fig0}
\end{figure}

 Next, we discuss the generic case in which some of the values of $w_\nu$ are complex.
 In this case, Eq.(\ref{sum-nu=0}) still holds and this implies that
 $\Re (w_\nu)_{\rm max} \ge  0$ and $\Re (w_\nu)_{\rm min} \le 0$,
 where $\Re (w_\nu)_{\rm max}$ and $\Re (w_\nu)_{\rm min}$
 indicate the maximum and minimum values of $\Re (w_\nu)$, respectively.
 (Here, ``$\Re $'' indicates real part.)
 In the complex plain, Eq.(\ref{con-WK2}) requires the existence of a point $a$,
 whose distances to all $w_\nu$ are shorter than the distance between $a$ and the point $(1,0)$.
 We use $Z_0$ to indicate a region in the complex plain, within which all the points $w_\nu$ lie.
 Then, Eq.(\ref{con-WK2}) means the existence of a circle centered at $a$,
 which covers the region $Z_0$ meanwhile does not contain the point $(1,0)$
 (see Fig.\ref{fig0}(a) for an example).

 We assume that the perturbation $V$ has real matrix elements in the unperturbed basis $|k\ra$.
 This requires that eigenvalues of $W_S$
 should appear in pairs with the form $(w_\nu, w_\nu^*)$.
 Thus, the region $Z_0$ is symmetric with respect to the horizontal axis.
 In the case that $\Re (w_\nu)_{\rm max} < 1$,
 one can always find a circle possessing the property discussed in the previous paragraph [see Fig.\ref{fig0}(a)].
 On the other hand, in the case that $\Re (w_\nu)_{\rm max} \ge 1$, noting that $\Re (w_\nu)_{\rm min} \le 0$,
 as illustrated in Fig.\ref{fig0}(b), such a circle does not exist.
 Therefore, finally, the condition (\ref{con-WK2}) reduces to the following one,
\be\label{CA}
\Re (w_\nu)_{\rm max} < 1.
\ee

 In what follows, for the simplicity in discussion,
 we do not consider the generic case with complex $w_\nu$.
 In other words, we only consider those operators $W_S$ that possess real spectra of $w_\nu$,
 for which the condition (\ref{wm-11}) is of relevance.

\subsection{Semiperturbative treatment}\label{sect-semiperturb}

 The main result of the above discussions,
 i.e., certain part of an EF being expanded in a convergent GBWPE,
 supplies a method for the study of structural properties of EFs.
 We call this method a \emph{semiperturbative} treatment,
 because each EF should be divided into
 two parts and only one part is expanded in a convergent GBWPE.

 In a semiperturbative treatment to a given perturbed state $|\alpha\ra$,
 one is often interested in the largest set of $S$,
 for which the condition (\ref{wm-11}) is satisfied.
 We use $S_{\rm PT}$ to denote such a set $S$
 and call it the \emph{perturbative} (PT) region of $|\alpha\ra$.
 The part of the state $|\alpha\ra$ lying in this region,
 namely, $|\alpha_P\ra = P_{S_{\rm PT}}|\alpha\ra$ is called the PT part of $|\alpha\ra$.
 Correspondingly, we call the complementary set of $S_{\rm PT}$,
 denoted by $S_{\rm NPT}$,
 the \emph{nonperturbative} (NPT) region of $|\alpha\ra$ and
 the corresponding part $|\alpha_Q\ra$ the NPT part of $|\alpha\ra$.

 Below, we discuss some generic features of the NPT region $S_{\rm NPT}$,
 with respect to the perturbation strength $\lambda$.
 When $|\lambda |$ is sufficient small,  $S_{\rm NPT}$ contains only one unperturbed state $|k\ra$,
 whose eigenvalue $E^0_k$ is the closest to $E_\alpha$.
 In fact, due to the linear dependence of the operator $W_S$ on $\lambda$,
 a sufficiently small $\lambda$ would
 guarantee validity of the condition (\ref{wm-11}).
 In this case, according to Eq.(\ref{con-K}) one may choose $a=0$ and,
 as a result, Eq.(\ref{alpha-ovs-K}) gives an ordinary perturbation expansion.

 When $|\lambda|$ increases beyond some value which is usually still small,
 the condition (\ref{wm-11}) will be violated for the set $S_{\rm NPT}$ discussed above.
 Sometimes, changing the state $|k\ra \in S_{\rm NPT}$ to some other
 unperturbed state may resume validity of the condition (\ref{wm-11}).
 In this case, the NPT region $S_{\rm NPT}$ still contains only one unperturbed state.
 But, generically, in order to satisfy the condition (\ref{wm-11}),
 the set $S_{\rm NPT}$ should be enlarged.

 To get a better understanding, let us substitute the operator $P_S$
 in Eq.(\ref{PQ}) into the expression of $W_S$ in Eq.(\ref{U}), which gives
\begin{gather}\label{WS-exp}
 W_S =   \sum_{|k\ra , |k'\ra \in S}   \frac{\la k'|\lambda  V|k\ra}{E_\alpha-E^0_k} |k'\ra \la k|.
\end{gather}
 From the rhs of Eq.(\ref{WS-exp}), it is seen that the value of $|w_{\rm max}|$
 is  usually decreased, when those unperturbed states that have large values
 of $ \left | \frac{\la k'|\lambda  V|k\ra}{E_\alpha-E^0_k}\right | $ are moved out of the set $S$.
 Therefore, loosely speaking, with the increase of $|\lambda|$,  the size of the NPT region should increase
 and that of  the PT region should decrease.

 %With $\lambda$ increasing further, the set $S_{\rm NPT}$ becomes larger and larger.
 %In this process, the continuity of the increase of $\lambda$ implies that it is
 %the condition (\ref{wm-11}) that is satisfied.
 %For this reason, in most applications, the type-I PT region is more useful
 %than the type-II region.

 {Finally, we mention some properties useful in the study of PT and NPT regions, which can be seen from
 discussions given above.}
 (i) If the condition (\ref{inequ}) is satisfied for a set $S$, then, usually
 it is also satisfied for subsets of $S$.
 (ii)  The condition (\ref{inequ}) is satisfied
 for a set $S$, if $S$ contains only those unperturbed states $|k\ra$ whose unperturbed energies
 are sufficiently far from $E_\alpha$.

\section{Semiclassical meaning of PT regions of EFs}\label{sect-NPT}

 In this section, for perturbed systems that possess classical counterparts,
 we discuss a semiclassical meaning of PT regions of EFs.
 Specifically, in Sec.\ref{sect-ps-exres-cond} we derive a phase-space presentation of the condition (\ref{wm-11})
 in the semiclassical limit, then, in Sec.\ref{sect-PT-cforbid} we show that
 PT regions  can be connected to classically-forbidden regions.

\subsection{Phase-space expression of the condition for PT region in the semiclassical limit}\label{sect-ps-exres-cond}

 We consider a perturbed system $H$ that has an $f$-dimensional classical counterpart
 and an unperturbed system $H_0$ whose classical counterpart is  an integrable system.
 In terms of action-angle variables, $H_0$ is a function of the action $\bI$ only,
 where ${\boldsymbol{I}}=({I_1},{I_2},\cdots,{I_f})$, and $H$ is written as
\be
H(\boldsymbol{I},\boldsymbol{\theta})=H_{0}(\boldsymbol{I})+\lambda V(\boldsymbol{I},\boldsymbol{\theta}).
\ee
 For the simplicity in discussion, we assume that $H_0$ is simply written as
\be
H_0=\boldsymbol{d}\cdot {\boldsymbol{I}} +c_0,
\ee
 where $\boldsymbol{d }$
 is a parameter vector, $\boldsymbol{d }=(d _{1},d _{2},\cdots,d _{f})$, and $c_0$ is a single parameter.
 We use $|\boldsymbol{m} \ra$, with an integer vector $\boldsymbol{m}=(m_1,m_2,\cdots,m_f)$,
 to indicate the unperturbed states $|k\ra$ discussed in previous sections.
 Here, they are eigenstates of the action operator ${\boldsymbol{I}}$,
 ${\boldsymbol I} |\boldsymbol{m}\ra = \boldsymbol {I_m}|\boldsymbol{m}\ra$, where
\begin{gather}\label{eq-Ik}
 \boldsymbol{I_m}=\boldsymbol m \hbar,
\end{gather}
 and $H_{0}|\boldsymbol{m}\rangle=E_{\boldsymbol{m}}^{0}|\boldsymbol{m}\rangle$
 with $E_{\boldsymbol{m}}^{0} = \hbar \boldsymbol{d}\cdot {\boldsymbol{m}}  +c_0$.

 To find a physical meaning of the PT region of a perturbed state $|\alpha\ra$,
 it is not convenient to study the operator $W_S$, which is not Hermitian.
 We are to go back to the condition in Eq.(\ref{conditionWsE})
 and study the Hermitian operator $A_n$ and the quantity $A_n^\psi$.
 We are to express $A_n^\psi$ in the coherent-state representation with coherent states
 denoted by $|c\rangle$.
 Below, we first discuss a one-dimensional system, then, generalize the results to be obtained
 to a generic $f$-dimensional system.

 In a one-dimensional system, a coherence state is written as follows in the unperturbed states $|m\rangle$
 \cite{QO},
\be\label{eq-alpha}
|c\rangle=\exp\left(\frac{-|c|^{2}}{2}\right)\sum_{m=0}^{\infty}\frac{c^{m}}{\sqrt{m!}}|m\rangle.
\ee
 In the coherent-state representation,
 the quantity $A_n^\psi$ in Eq.(\ref{An-psi}) is written as \cite{Husimi,HusimiP}
\be
A_n^\psi =\frac{1}{\pi}\int dc^{2}\rho_\psi(c^{{*}},c)\exp\left(\overleftarrow
{\frac{\partial}{\partial c}}\overrightarrow{\frac{\partial}{\partial c^{{*}}}}\right) A_n(c^{{*}},c),
\ee
 where $\rho_\psi(c^*,c)=\langle c |\psi\ra \la \psi |c \rangle$ and
\begin{gather}
A_{n}(c^{*},c)=\langle c|A_{n}|c\rangle \nonumber \\
=\langle c|\left (P_{S}\frac{\lambda V\frac{1}{E-H}-a}{1-a}P_{S} \right)^{n} \left ( P_{S}\frac{\frac{1}{E-H}\lambda V-a}{1-a}P_{S} \right )^{n}|c\rangle .\label{An-aa*}
\end{gather}

 Suppose that the system has an effective Planck constant $\hbar$,
 which may go to zero as the semiclassical limit.
 The complex variables $(c,c^{{*}})$ can be written in the following way,
 in terms of $\hbar $ and a pair of real variables $({J},{\theta})$,
\be\label{eq-aa}
c=\sqrt{\frac{J}{\hbar}}\exp\left(-\frac{i\theta}{2}\right),
 \quad c^{{*}}=\sqrt{\frac{J}{\hbar}}\exp\left(\frac{i\theta}{2}\right).
\ee
 Straightforward derivation gives that
\be \notag
\overleftarrow{\frac{\partial}{\partial c}}\overrightarrow
{\frac{\partial}{\partial c^{{*}}}}=\hbar\left(\sqrt{J}\overleftarrow{\frac{\partial}{\partial J}}
+i\sqrt{\frac{1}{J}}\overleftarrow{\frac{\partial}{\partial\theta}}\right)
\left(\sqrt{J}\overrightarrow{\frac{\partial}{\partial J}}-i\sqrt{\frac{1}{J}}
\overrightarrow{\frac{\partial}{\partial\theta}}\right).
\ee
 Thus, in the semiclassical limit of $\hbar \to 0$, one gets that
 $\exp(\overleftarrow{\frac{\partial}{\partial c}}\overrightarrow{\frac{\partial}{\partial c^{{*}}}}) =1$
 and, as a result,
\be\label{trRW}
 A_n^\psi =\frac{1}{\pi}\int dc^{2}\rho_\psi(c^{{*}},c) A_n(c^{{*}},c).
\ee
It is known that, in the semiclassical limit, one has
\be\label{eq-Cpro}
\langle c|G_1 G_2|c\rangle= \langle c| G_1 |c\rangle \langle c| G_2 | c \rangle,
\ee
 where $G_1$ and $G_2$ are two arbitrary operators.
 Hence, once the  expressions of $\langle c| P_S | c \rangle $ and of $\langle c|\frac{1}{E-H}\lambda V|c\rangle$ are known, one can compute the quantity $A_n(c^{{*}},c)$ in Eq.(\ref{An-aa*}).

 For an arbitrary set $S$, $P_S$ is written as
\begin{gather}\label{PS-int}
 P_S=\sum_{{|m\ra}\in S}|{m}\rangle \langle {m}|.
\end{gather}
 Let us study the overlap $f_{c}(m)$,
\be
 f_{c}(m) = |\langle m|c\rangle|^2.
\ee
 Making use of Eqs.(\ref{eq-alpha}) and (\ref{eq-aa}), one finds that
\begin{align}\label{fJtheta}
 |f_{c}(m)|^2 %& =\frac{(\frac{J}{\hbar})^{m}\exp(-\frac{J}{\hbar})}{m!}
                   & = \exp\left[m\ln\left(\frac{J}{\hbar}\right)-\frac{J}{\hbar}-\ln m!\right].
\end{align}
 For a fixed value of $I_m$, when $\hbar$ is sufficient small,
 one has $m=\frac{I_m}{\hbar}\gg 1$, such that
\be\label{m!}
 m! \simeq \sqrt{2m\pi}\left(\frac{m}{e}\right)^{m}.
\ee

 To go to the semiclassical limit finally, it is convenient to use $I_m = m \hbar$ to
 replace $m$ in the above expressions.
 For brevity, we drop the subscript $m$ of $I_m$ in what follows.
 Then, substituting Eq.(\ref{m!}) into Eq.(\ref{fJtheta}), one finds that
\be\label{eq-fc}
f_{c}(I)=\exp\left(\frac{I(\ln J-\ln I)}{\hbar}-\frac{J-I}{\hbar}-\frac{1}{2}\ln\frac{2\pi I}{\hbar}\right).
\ee
To study properties of $f_{c}(I)$, one can written it in a more compact form as
\be
f_{c}(I)=\sqrt{\frac{\hbar}{2\pi I}}e^{F(I)},
\ee
where
\be
F(I)=\frac{I(\ln J-\ln I)}{\hbar}-\frac{J-I}{\hbar}.
\ee
As the prefactor $\sqrt{\frac{\hbar}{2\pi I}}$ varies slowly with $I$,
one can concentrate on the function $F(I)$.
 We note that
\be\label{eq-df}
\frac{dF}{dI}=\frac{\ln J-\ln I}{\hbar},\quad \frac{d^{2}F}{dI^{2}}=-\frac{1}{I\hbar}.
\ee
 Hence, $F(I)$ has a very sharp peak  at $I=J$ for small $\hbar$,
 with a width proportion to $\sqrt{\hbar}$.
 The narrowness of the peak implies that $I$ in the prefactor can be taken as $J$.
 Moreover, in the neighbourhood of $J$, $f_c(I)$ can be written as
\be\label{eq-FJX1}
f_{c}(I) = \sqrt{\frac{\hbar}{2\pi J}}\exp\left(-\frac{(I-J)^{2}}{2J\hbar}\right).
\ee
 Then, noting that
\be
\delta(x)=\lim_{a\rightarrow0}\frac{1}{a\sqrt{\pi}}\exp(-\frac{x^{2}}{a^{2}}),
\ee
 in the semiclassical limit $\hbar\rightarrow 0$, $f_{c}(I)$ approach a $\delta$ function, that is,
\be\label{fcI-1}
f_{c}(I) = \hbar\delta(I-J).
\ee

 It is straightforward to generalize the above discussions to the generic case
 of an $f$-dimension system, in which the coherent states are written as
\be
|c\rangle=\exp\left(-\frac{1}{2}\sum_{i=1}^{f}|c_{i}|^{2}\right)
\prod_{{i}=1}^{f}\left[\sum_{m_{i}=1}^{\infty}\frac{(c_{i})^{m_{i}}}{\sqrt{m_{i}!}}|m_{i}\rangle\right].
\ee
 Similar to Eq.(\ref{fcI-1}), the overlap $f_{c}(\bI) = |\langle \bm|c\rangle|^2$
 has the following expression in the
 semiclassical limit,
\be\label{fcI-f}
f_{c}(\bI) = \hbar^f \delta^f(\bI-\bJ).
\ee
This gives that
\be\label{eq-Ps}
\langle c| P_{S}|c \rangle=\begin{cases}
1, & |\bJ\ra \in S; \\
0, & |\bJ\ra \notin S.
\end{cases}
\ee

Note that in the semiclassical limit one has
\be\label{WC}
\langle c|\frac{1}{E-H}\lambda V|c\rangle=\frac{\lambda V(\boldsymbol{J},\boldsymbol{\theta})}{E-H_{0}(\boldsymbol{J})}\equiv W_{c}(\boldsymbol{J},\boldsymbol{\theta}).
\ee
 Making use of Eqs.(\ref{eq-Cpro}), (\ref{eq-Ps}) and (\ref{WC}),
 one finds that $A_n(c,c^*)$ in Eq.(\ref{An-aa*}) has the following expression,
\be\label{eq-WSF}
A_{n}(c,c^*)=\begin{cases}
(W_{a}(\boldsymbol{J},\boldsymbol{\theta}))^{2n}, & \boldsymbol{J}\in S;\\
0, & \boldsymbol{J}\notin S.
\end{cases}
\ee
where,
\be\label{Hws}
W_{a}(\boldsymbol{J},\boldsymbol{\theta})=\frac{W_c(\boldsymbol{J},\boldsymbol{\theta})-a}{1-a}.
\ee
 Substituting Eq.(\ref{eq-WSF}) into Eq.(\ref{trRW}) for $A^\psi _n$,
 it is seen that the condition (\ref{conditionWsE}) is equivalent to the requirement
 that there exists some parameter $a$, such that
\be\label{eq-N}
\left|\frac{W_c(\boldsymbol{J},\boldsymbol{\theta})-a}{1-a}\right|<1, \quad
\forall(\boldsymbol{J},\boldsymbol{\theta})\text{ with }|\boldsymbol{J}\ra \in S.
\ee

\subsection{PT regions as classically forbidden regions}\label{sect-PT-cforbid}

 In order to see more clearly the geometric meaning of the condition (\ref{eq-N}),
 let us divide the phase space into regions, denoted by $\gamma(\boldsymbol{I})$,
 each possessing a definite value of the action variable $\bI$, that is,
\begin{gather}\label{}
 \text{phase space} = \bigcup_{\bI} \gamma(\bI),
 \\ \gamma(\boldsymbol{I})=\{(\boldsymbol{I},\boldsymbol{\theta}):
 \boldsymbol{\theta} \in[0,2\pi]^{f}\}.
\end{gather}
 For brevity in presentation, we may use $\gamma(\bI)$ as a variable of a function in a relation.
 When we do this, we mean that the relation holds for all points in the region $\gamma(\bI)$.
 For example, $F(\gamma(\boldsymbol{I}))<1$ means that $F(\boldsymbol{I},\boldsymbol{\theta})<1$ hold
 for all points $(\boldsymbol{I},\boldsymbol{\theta})\in\gamma(\boldsymbol{I})$.

 We use $\gamma_S $ to denote the region in the phase space that corresponds to a
 given set $S$ of unperturbed states $|\bm\ra$,
\be
 \gamma_S = \bigcup_{|\bm\ra \in S} \gamma(\boldsymbol{I}_{\bm}).
\ee
 In the limit $\hbar \to 0$, $\bI_{\bm}$ in Eq.(\ref{eq-Ik}) becomes continuous.
 Then, the condition (\ref{eq-N}) can be reexpressed as follows,
\be\label{eq-N-gamma}
\left|\frac{W_c(\gamma(\boldsymbol{I}))-a}{1-a}\right|<1, \quad
\forall\gamma(\boldsymbol{I})\in\gamma_{S}.
\ee

 Then, following arguments similar to those given previously between Eq.(\ref{con-WK2})
 and Eq.(\ref{wm-11}), we find that the condition (\ref{eq-N-gamma}) reduces to that
 $W_{c}(\gamma(\boldsymbol{I}))<1$ for all $\gamma(\boldsymbol{I})\in\gamma_{S}$.
 In this derivation, we have used the property that, for each given $\bJ$,
 the classical quantity $V(\boldsymbol{J},\boldsymbol{\theta})$ can be both positive and negative;
 this property is a consequence of the fact of $V$ possessing zero diagonal elements
 in the unperturbed basis, which in the semiclassical limit implies that
$\int_{0}^{2\pi}V(\boldsymbol{J},\boldsymbol{\theta})d\boldsymbol{\theta}=0$.

 In fact, the requirement of $W_{c}(\gamma(\boldsymbol{I}))<1$ is equivalent to
 that of $W_{c}(\gamma(\boldsymbol{I})) \ne 1$.
 To establish this point, we first note that if $W_{c}(\gamma(\boldsymbol{I}))<1$,
 then, $W_{c}(\gamma(\boldsymbol{I})) \ne 1$.
 Next, suppose that $W_{c}(\gamma(\boldsymbol{I})) \ne 1$,
 which means that $W_{c}(\boldsymbol{I},\boldsymbol{\theta})\ne 1$  for all $\boldsymbol{\theta}$.
 Since $W_c(\boldsymbol{I},\boldsymbol{\theta})$ is a continuous function of $\boldsymbol{\theta}$,
 $W_{c}(\gamma(\boldsymbol{I})) \ne 1$ implies that either
 $W_c(\boldsymbol{I},\boldsymbol{\theta})>1$ for all $\boldsymbol{\theta}$,
 or $W_c(\boldsymbol{I},\boldsymbol{\theta})<1$ for all $\boldsymbol{\theta}$.
 As discussed above,  for each given $\bI$,
 $V(\boldsymbol{I},\boldsymbol{\theta})$ must be positive for some $\boldsymbol{\theta}$
 and be negative for some other $\boldsymbol{\theta}$.
 This excludes the possibility that
 $W_c(\boldsymbol{I},\boldsymbol{\theta})>1$ for all $\boldsymbol{\theta}$.
 Hence, $W_c(\boldsymbol{I},\boldsymbol{\theta})$ must be less than $1$ for all $\boldsymbol{\theta}$,
 i.e., $W_{c}(\gamma(\boldsymbol{I}))<1$.

 Thus, the condition (\ref{eq-N-gamma}) is written in the following form,
\be\label{eq-WCG}
H(\gamma(\boldsymbol{I}))\neq E,\quad \forall\gamma(\boldsymbol{I})\in\gamma_{S}.
\ee
 The physical meaning of $H(\gamma(\boldsymbol{I})) \ne E$ is quite clear,
 that is, all the points in the region $\gamma(\boldsymbol{I})$ are
 classically forbidden for the energy $E$.
 We call such a region $\gamma(\bI)$ a \emph{classically-forbidden region}.
 Thus, from the above discussions, we reach the following conclusion as the main result of this paper.
\begin{itemize}
  \item In the semiclassical limit, the PT region of
  a perturbed state $|\alpha\ra$ correspond to the maximum classically forbidden
  region of $\gamma(\bI)$ in the phase space.
\end{itemize}

\begin{figure}
\includegraphics[width=1\linewidth]{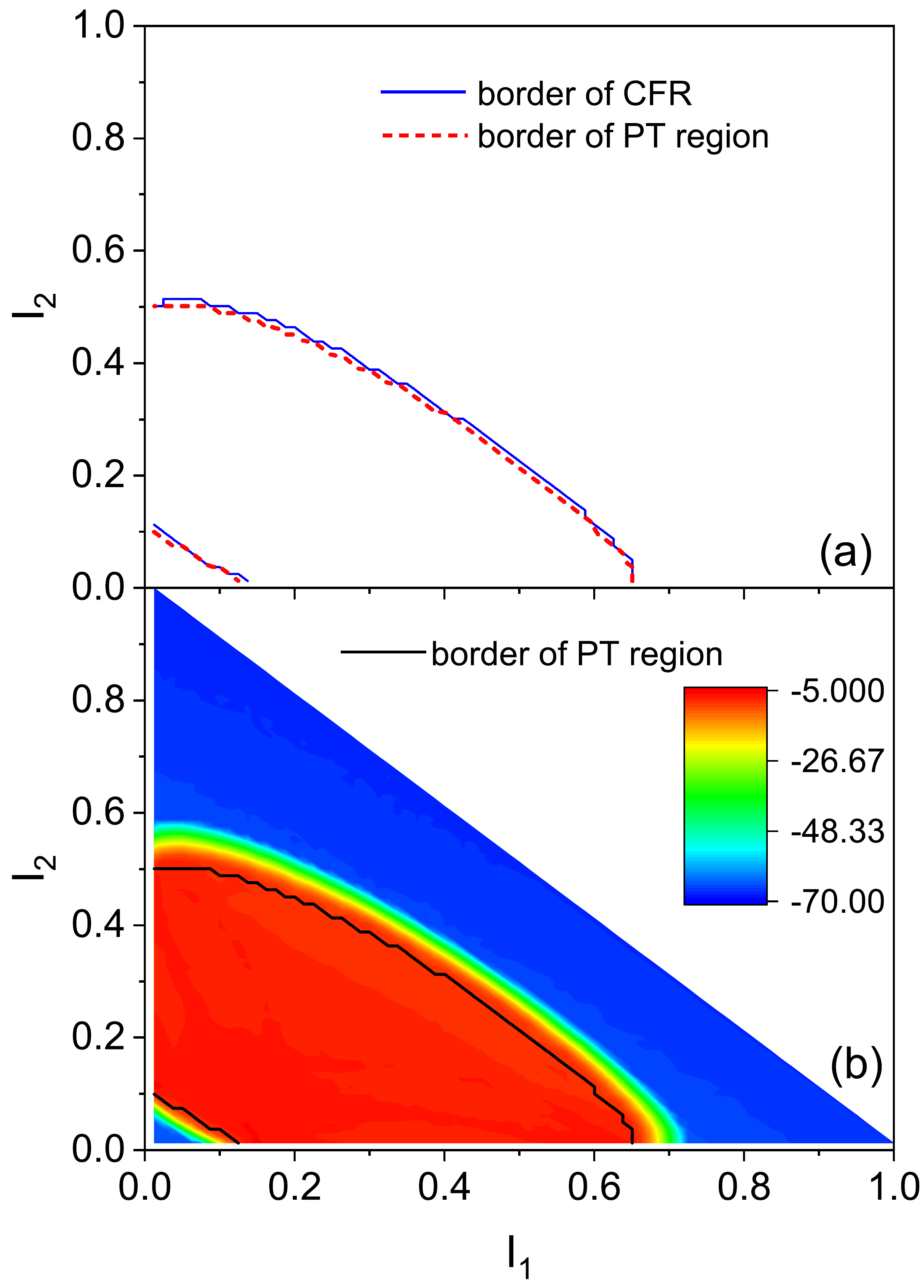}\caption{
 (a) Comparison of the border of the PT region of a perturbed state $|\alpha\ra$
 with an energy $E$ in LMG model,
 indicated by the dashed lines (in read), and that of the classically forbidden region(CFR)
 for the same energy, indicated by the solid lines (in blue).
 Parameters: $\Omega=1999$,  $\lambda=0.5$, and $E \simeq 13.4$.
(b) The averaged shape of the EF, $\langle|C_{\alpha \boldsymbol{m}}|^{2}\rangle$ in Eq.(\ref{eq-AEF}),
  in the logarithmic scale showed by means of color.
  Five neighboring EFs were used in the computation.
 The PT border is indicated by the black lines in the red region.
}\label{fig-NPTQC-LMG}
\end{figure}

\begin{figure}
\includegraphics[width=1\linewidth]{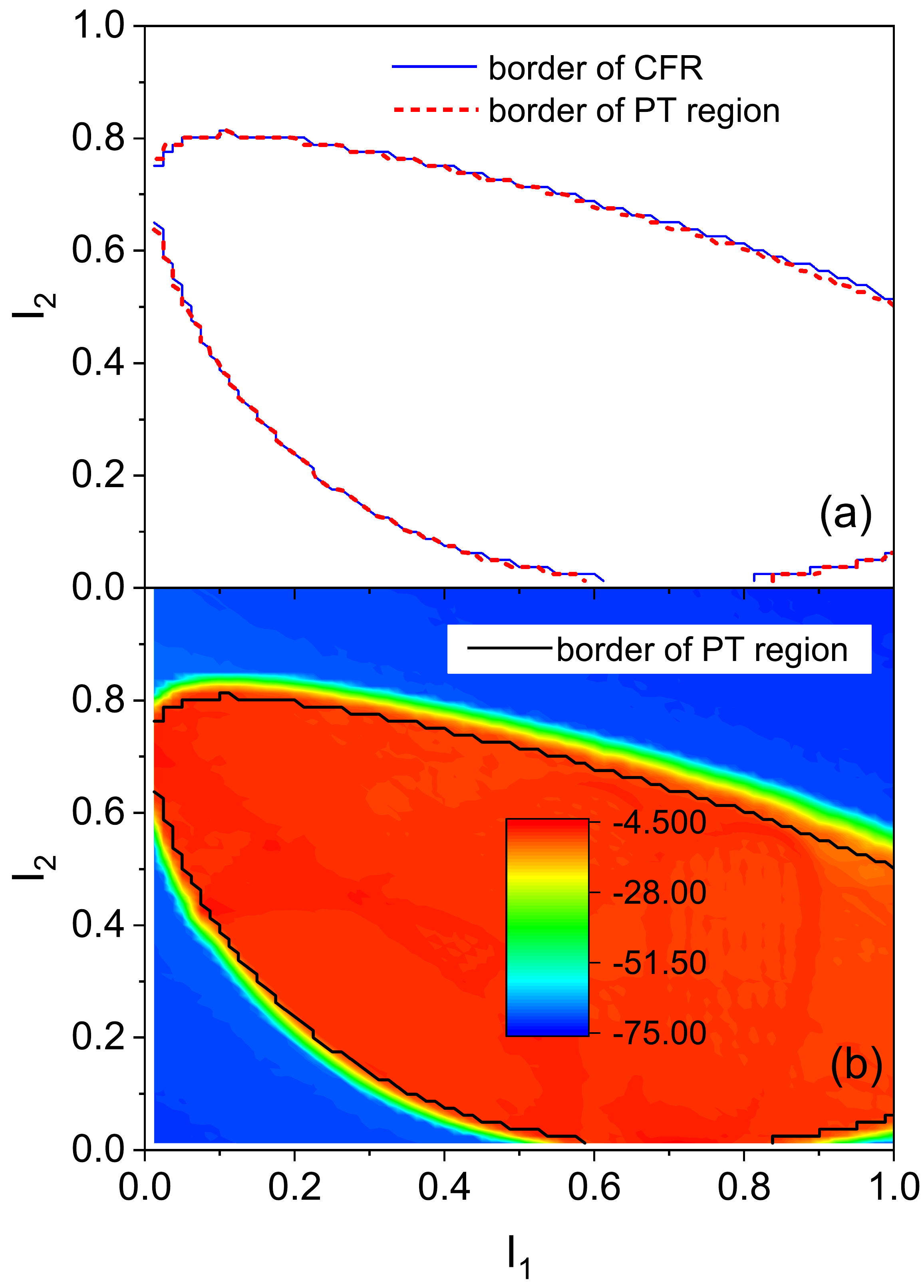}
\caption{Similar to Fig.\ref{fig-NPTQC-LMG}, but for the Dicke model with $N=1999$,  $\lambda=0.4$,
 and $E \simeq 0.21$.
}\label{fig-NPTQC-Dicke}
\end{figure}

\section{Numerical simulations }\label{app-models}

 In this section, we discuss numerical tests for the main result given above.
 We do this in three models, a Lipkin-Meshkov-Glick (LMG) model, a Dicke model,
 and a three-site Bose-Hubbard model.

The first model we employ is a three-orbital LMG model \cite{LMG}.
 This model is composed of $\Omega$ particles, occupying three energy levels labeled by $r=0,1,2$, each with
 $\Omega$-degeneracy.
 Here, we are interested in the collective motion of this model, for which the dimension of the
 Hilbert space is $\frac 12 (\Omega+1)(\Omega +2)$.
 We use $\epsilon_{r}$ to denote the energy of the $r$-th level
 and, for brevity, we set $\epsilon_{0}=0$.
 The Hamiltonian of the model, in the form in Eq.(\ref{H}), is written as \cite{pre-98}
\begin{gather}
 H_{0}=\epsilon_{1}K_{11}+\epsilon_{2}K_{22}  \\
V=\sum_{t=1}^{4}\mu_{t}V^{(t)}.
\end{gather}
 Here, $K_{rr} $ represents the particle number operator for the level $r$ and
\begin{eqnarray}
V^{(1)}=K_{10}K_{10}+K_{01}K_{01},\ V^{(2)}=K_{20}K_{20}+K_{02}K_{02},\nonumber \\
 V^{(3)}=K_{21}K_{20}+K_{02}K_{12},\ V^{(4)}=K_{12}K_{10}+K_{01}K_{21}, \ \
\end{eqnarray}
 where $K_{rs}$ with $r\neq s$ indicate particle raising and lowering operators.
 The unperturbed states are written as $|m_1 m_2\ra$, for which
 $H_0 |m_1 m_2\ra = \epsilon_{1}m_1 +\epsilon_{2}m_2$.
 This model has an effective Planck constant $\hbar_{\rm eff}$,
\be\label{eq-hbarL}
\hbar_{\rm eff}=\frac{1}{\Omega}.
\ee
 Detailed properties of this model and of its classical counterpart can be found
 in Refs.\cite{pre-98,Meredith98}.

 Numerically, we found that, in agreement with the main result given in the previous section,
 borders of the PT regions of
 perturbed states are close to the corresponding classically-forbidden regions,
 when $\Omega$ is large.
 One example is given in Fig.\ref{fig-NPTQC-LMG}(a), in which
 the classical counterpart of the model
 is fixed with the requirement that $\epsilon_{1} \Omega= 44.1, \epsilon_{2} \Omega= 64.5,
\mu_{1} \Omega^2 = 62.1, \mu_{2} \Omega^2 = 70.2, \mu_{3} \Omega^2= 76.5$, and $\mu_{4} \Omega^2= 65.7$.
 We also plot the averaged shape of neighboring EFs (Fig.\ref{fig-NPTQC-LMG}(b)), 
 denoted by $\langle|C_{\alpha\boldsymbol{m}}|^{2}\rangle$,
\be\label{eq-AEF}
\langle|C_{\alpha\boldsymbol{m}}|^{2}\rangle=\frac{1}{N_{E_{\alpha}}}
\sum_{\alpha'}|C_{\alpha'\boldsymbol{m}}|^{2}
\delta_{\epsilon}(E_{\alpha'}-E_{\alpha}),
\ee
where
\be\label{eq-NEA}
N_{E_{\alpha}}=\sum_{\alpha{'}}\delta_{\epsilon}(E_{\alpha{'}}-E_{\alpha}).
\ee
Here $\delta_\epsilon (E)$ is a coarse-grained $\delta$-function,
\be
\delta_{\epsilon}(E)=\begin{cases}
\frac{1}{\epsilon} & E\in[-\frac{\epsilon}{2},\frac{\epsilon}{2}],\\
0 & {\rm otherwise},
\end{cases}
\ee
 where $\epsilon$ is a small parameter.

The second model is a single-mode Dicke model\cite{Dicke,Emary03},
which describes the interaction between a single bosonic mode and
a collection of $N$ two-level atoms. The system can be described
with the collective operator ${\bf {J}}$ for the $N$
atoms, with
\begin{equation}
{J}_{z}\equiv\sum_{i=1}^{N}{s}_{z}^{(i)},\ \ {J}_{\pm}\equiv\sum_{i=1}^{N}{s}_{\pm}^{(i)},
\end{equation}
where ${s}_{x(y,z)}^{(i)}$ are Pauli matrices divided by $2$
for the $i$-th atom.
The Dicke Hamiltonian is written as~\cite{Emary03}
\begin{equation}
H=\omega_{0}J_{z}+\omega a^{\dagger}a+\frac{\lambda}{\sqrt{N}}\mu(a^{\dagger}+a)(J_{+}+J_{-}),
\end{equation}
 {which can also be written in the form of $H=H_0 + \lambda V$.}
The operator ${J}_{z}$ and ${J}_{\pm}$ obey the usual commutation
relations for the angular momentum,
\begin{equation}
[J_{z},J_{\pm}]=\pm J_{\pm},\ \ [J_{+},J_{-}]=2J_{z}.
\end{equation}

The Hilbert space of this model is spanned by vectors $|j,m\rangle$ with $m=-j,-j+1,\cdots,j-1,j$,
which are known as the Dicke states and are eigenstates of $\boldsymbol{J}^{2}$
and $J_{z}$, with $J_{z}|j,m\rangle=m|j,m\rangle$ and $\boldsymbol{J}^{2}|j,m\rangle=j(j+1)|j,m\rangle$.
 Below, we take $j$ as its maximal value, namely, $j=N/2$;
 it is a constant of motion, since $[\boldsymbol{J}^{2},H]=0$.
Another conserved observable in the Dicke model is the parity $\Pi$, which
is given by $\Pi=\exp(i\pi\hat{N})$,
 where $\hat{N}=a^{\dagger}a+J_{z}+j$ is an operator for the ``excitation number'',
 counting the total number of excitation quanta in the system. In
our numerical study, we consider the subspace with $\Pi=+1$.

Making use of the Holstein-Primakoff representation of the angular
momentum operators,
\begin{gather}
J_{+}=b^{\dagger}\sqrt{2j-b^{\dagger}b},\ \ \ J_{-}=\sqrt{2j-b^{\dagger}b} \ b,\nonumber \\
J_{z}=(b^{\dagger}b-j),\label{eq-J}
\end{gather}
where $b$ and $b^\dag$ are bosonic operators satisfying $[b,b^{\dagger}]=1$,
the Hamiltonian can be further written in the following form,
\begin{gather}\label{}\notag
 H=\omega_{0}(b^{\dagger}b-j)+\omega a^{\dagger}a
 \\ +\lambda\mu(a^{\dagger}+a) \left(b^{\dagger}\sqrt{1-\frac{b^{\dagger}b}{2j}}
 +\sqrt{1-\frac{b^{\dagger}b}{2j}}b\right).
\end{gather}
 From this Hamiltonian, a classical counterpart can be obtained in a direct way.

\begin{figure}
\includegraphics[width=1\linewidth]{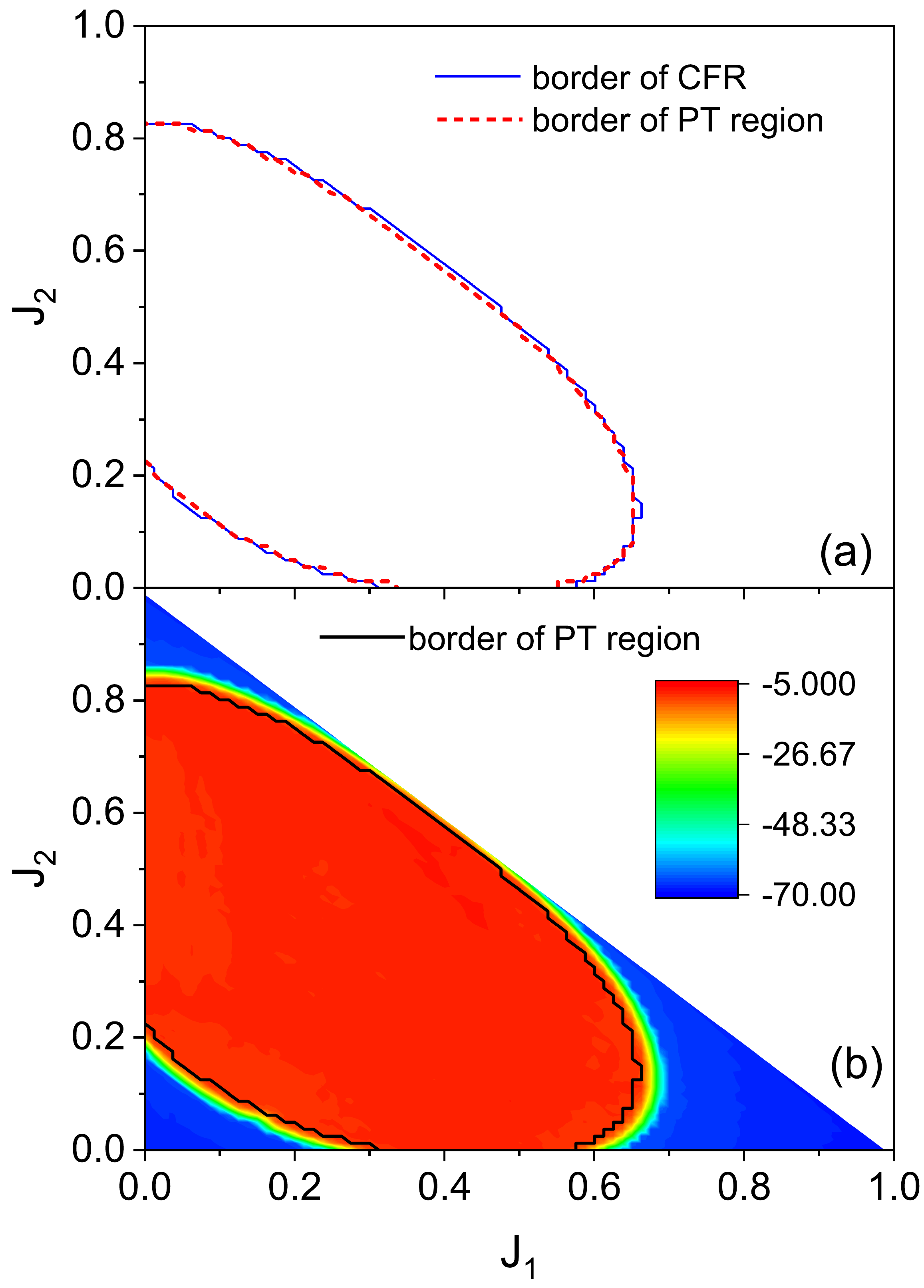}
\caption{Similar to Fig.\ref{fig-NPTQC-LMG}, but for the three-site
Bose-Hubbard model model with $N=1999$, $h=0.1$, $\lambda=0.1$, and $E \simeq 0.2$.
}\label{fig-NPTQC-BH}
\end{figure}

 We write the Fock states related to $a^\dag$ and $b^\dag$ as $|n_{a}\rangle$
and $|n_{b}\rangle$, respectively,  for which
\begin{equation}
a^{\dagger}a|n_{a}\rangle=n_{a}|n_{a}\rangle,\ \ \ b^{\dagger}b|n_{b}\rangle=n_{b}|n_{b}\rangle.
\end{equation}
According to Eq.(\ref{eq-J}), $n_{b}$ should be truncated at $(n_{b})_{max}=N$.
 In numerical simulations, we set $(n_{a})_{max}=N$.
 Other parameters are $\omega_0=\omega=1/N$ and $\mu=1/N$.
 This model has an effective Planck constant
\be\label{eff-h-Dicke}
\hbar_{\rm eff}=\frac{1}{N}.
\ee
 In the Dicke model, we also numerically found closeness of the PT regions to classically
 forbidden regions (see Fig.\ref{fig-NPTQC-Dicke} for an example).

 The third model is a three-site Bose-Hubbard (BH) model \cite{BHM},
 which describes interacting bosons on a lattice.
 It also possesses a classical counterpart \cite{Engl}.
 This model can be realized experimentally using ultracold quantum gases in optical lattices \cite{BHexp}.
 It undergoes a transition from a superfluid phase
 to an insulator phase as the strength of the potential is increased beyond some value \cite{QPT}.

 The Hamiltonian of the three-site Bose-Hubbard model is written as
 $H=H_0 +\lambda V$, where
\begin{gather}
 H_0=\sum_{j=1}^{3}\left[\frac{U}{2}n_{j}(n_{j}-1)+h\mu_{j}n_{j}\right], \nonumber \\
 V=\sum_{j=1}^{2}\left[-J(a_{j}^{\dagger}a_{j+1}+a_{j+1}^{\dagger}a_{j})\right].
\end{gather}
 Here, $n_j=a^{\dagger}_j a_j$ indicates the particle number operator at a site $j$.
 The total number of particles $N=\sum_{j=1}^L n_j$ is a conserved quantity.
 We use the open boundary condition in our numerical simulations.
 For each given value of $\lambda$, the parameters are chosen as $J=\frac{1}{N}$, $U=\frac{1}{N^2}$, $h=\frac{1}{N}$,
 $\mu_1=0.1$, $\mu_2=0$, and $\mu_3=-0.1$.
 The semiclassical limit of this model can be obtained,
 with an effective plank constant $\hbar_{\rm eff}=\frac{1}{N}$ going to zero
 in the limit $N\rightarrow \infty$ \cite{Engl}, with the transformation
\be
a_{j}\rightarrow\sqrt{NI_{j}}\exp(i\phi_{j}),\ a_{j}^{\dagger}\rightarrow\sqrt{NI_{j}}\exp(-i\phi_{j}).
\ee
 This gives the following classical Hamiltonian,
\begin{gather}
H_{0}=\frac{1}{2}(I_{1}^{2}+I_{2}^{2}+I_{3}^{2})+\mu_{1}I_{1}+\mu_{2}I_{2}+\mu_{3}I_{3}, \\
V=2\sqrt{I_{1}I_{2}}\cos(\phi_{1}-\phi_{2})+2\sqrt{I_{2}I_{3}}\cos(\phi_{2}-\phi_{3}).
\end{gather}
As the total action $I=I_1+I_2+I_3=1$ is a conserved quantity, the degrees of freedom
can be reduced from three to two \cite{BHC}, then, one gets
\begin{gather}\nonumber
H_{0}=\frac{1}{2}\left[J_{1}^{2}+J_{2}^{2}+(1-J_{1}-J_{2})^{2}\right] \\
+\mu_{1}J_{1}+\mu_{3}J_{2}+\mu_{2}(1-J_{1}-J_{2}),  \\
 \nonumber V=2\sqrt{J_{1}(1-J_{1}-J_{2})}\cos(\theta_{1})
 \\ +2\sqrt{J_{2}(1-J_{1}-J_{2})}\cos(\theta_{2}).
\end{gather}

 Similar to the two models discussed above, we got numerical results
 in agreement with the analytical prediction for closeness of the PT regions
 and classically forbidden regions (Fig.\ref{fig-NPTQC-BH} for an example).

\section{Conclusions and discussions}\label{sect-Conclusion}

 In this paper, we have shown that PT regions of perturbed states
 should, in the semiclassical limit, coincide with classically-forbidden regions
 for the corresponding energies.
 This implies that components of EFs in unperturbed states lying in classically-forbidden
 regions can be expanded in convergent perturbation expansions (GBWPE),
 making use of their components in classically-allowed regions.
 This supplies an analytical method, which should be useful in the
 study of EF components in classically-forbidden regions.

 Previous studies show that the GBWPE is indeed a useful tool
 in the study of structural properties of EFs.
 In fact, it can be used in the study of exponential-type decay of long tails of EFs
 \cite{pre-98,pre00}.
 The criterion for its convergence may be used
 to estimate the location of main bodies of EFs \cite{pre00,pre02-LMG,pre01-trunc,EFchaos-WW}.
 It is also useful in explaining
 a localization feature numerically observed in the Wigner-band random-matrix model \cite{pre02-WBRM},
 and it supplies an alternative way of understanding the Anderson localization \cite{ctp01-ratio}.
 Besides, it supplies an interesting approach to the tunneling effect \cite{cpl04-config,cpl05}.

 The above result of this paper for EF components in classically-forbidden regions,
 if combined with the well-known fact that the semiclassical theory is applicable to EF
 components in classically-allowed regions,
 supplies a useful framework for the study of properties of EFs in the whole energy region.
 It is our hope that this framework may be useful in future investigations
 in various topics, such as thermalisation processes in chaotic systems.

\acknowledgements

% The author is grateful to Yan Gu for valuable discussions and suggestions.
 This work was partially supported by the Natural Science Foundation of China under Grant
 Nos.~11275179, 11535011, and 11775210.

\end{document}